\documentclass[12pt]{article}
\usepackage{epsfig}
\usepackage{axodraw}

\newcommand{\be}{\begin{equation}}
\newcommand{\ee}{\end{equation}}
\newcommand{\bq}{\begin{eqnarray}}
\newcommand{\eq}{\end{eqnarray}}
\newcommand{\one}{1 \!\! 1}
\newcommand{\D}{\mathrm{d}}
\newcommand{\E}{\mathrm{e}}
\newcommand{\I}{\mathrm{i}}

\def\lsim{\mathrel{\rlap{\lower4pt\hbox{\hskip1pt$\sim$}}\raise1pt\hbox{$<$}}}
\def\gsim{\mathrel{\rlap{\lower4pt\hbox{\hskip1pt$\sim$}}\raise1pt\hbox{$>$}}}
\def\Vec#1{\mathpalette{\VVec}{#1}}                  
\def\VVec#1#2{\mbox{\boldmath$#1#2$\unboldmath}}
\def\nostrocostruttino#1\over#2{\mathrel{\mathop{\kern 0pt \rlap
{\hbox{$#1$}}} \hbox{\kern-.135em $#2$}}}

\newlength{\dhatheight}

\newcommand{\Strut}{\rule[-1.7ex]{0pt}{4.7ex}}        
\newcommand{\STRUT}{\rule[-2.1ex]{0pt}{5.5ex}}        
\def\anti#1{\mathpalette{\@anti}{#1}#1}
\def\@anti#1#2{\sbox0{$#1#2$}
  \makebox[0pt][l]{$#1\kern.30\ht0\overline{\kern-.35\ht0\phantom{#2}}$}}

\begin{document}

\pagestyle{empty}

\null
\vspace{2cm}

\begin{center}
{\bf \Large Double hadron lepto-production }
\vskip 6pt
{\bf \Large in the current and target fragmentation regions}
\end{center}

\vspace{1cm}

\begin{center}

{\large M.~Anselmino$^{a}$, V.~Barone$^b$, A.~Kotzinian$^{a,c}$}

\vspace{0.2cm}

$^a${\it Dipartimento di Fisica Teorica, Universit{\`a}
di Torino; \\
INFN, Sezione di Torino, 10125 Torino, Italy}

$^b${\it Di.S.T.A., Universit{\`a} del Piemonte
Orientale ``A. Avogadro''; \\
INFN, Gruppo Collegato di Alessandria,  15121 Alessandria, Italy}

$^c${\it Yerevan Physics Institute, 375036 Yerevan, Armenia}
\end{center}

\vspace{2cm}
\begin{center}
{\large ABSTRACT}
\end{center}
\vspace{0.5cm}
\noindent 
We study the inclusive production of two hadrons in deep inelastic processes, 
$\ell \, N \to \ell \, h_1 \, h_2 \, X$, with $h_1$ in the current 
fragmentation region (CFR) and $h_2$ in the target fragmentation region 
(TFR). Assuming a factorized scheme, the recently introduced polarized and transverse momentum dependent fracture functions couple to the transverse 
momentum dependent fragmentation functions. This allows the full exploration 
of the fracture functions for transversely polarized quarks. Some 
particular cases are considered.   
\vspace{0.5cm}


\newpage

\pagestyle{plain}

\section{Introduction}

In a recent paper \cite{Anselmino:2011ss} we have introduced the formalism 
of polarized and transverse momentum dependent fracture functions to describe
semi-inclusive deep inelastic scattering in the target fragmentation region.
We have shown that, at leading order of QCD, considering only the production 
of spinless or unpolarized hadrons, there are 16 independent fracture 
functions, which describe the conditional probabilities of finding 
unpolarized or polarized quarks inside unpolarized or polarized nucleons 
fragmenting into the final observed hadron. We have also given explicit sum 
rules which, upon integration over the momentum of the final hadron in the 
TFR, relate the fracture functions to the usual transverse momentum dependent 
distribution functions (TMDs).     
   
The 16 fracture functions can be divided into three classes, referring 
respectively to unpolarized (4), longitudinally polarized (4) and 
transversely polarized (8) quarks. The first 8 can be accessed in 
single-hadron or single-hadron + jet production: explicit expressions of 
the corresponding cross sections have been derived and interesting azimuthal dependences, which have to be compared with those found for single hadrons 
produced in the CFR, have been discussed \cite{Anselmino:2011ss}.             

The 8 fracture functions related to transversely polarized quarks are 
chiral-odd quantities and can only appear in physical observables 
containing another chiral-odd function. This can be achieved by considering 
the combined production of two hadrons, one in the TFR and one in the CFR; 
the production of the latter is described by transverse momentum dependent 
fragmentation functions and the Collins mechanism provides the necessary 
chiral-odd quantity. We develop here the full formalism for such a double 
hadron lepto-production, following and completing the work of 
Ref. \cite{Anselmino:2011ss}.   

\section{Double hadron lepto-production}


Let us start from a general two-particle inclusive lepto-production 
process
\[
l (\ell) + N(P) \rightarrow l (\ell') + h_1 (P_1) + h_2 (P_2) + X (P_X) \,. 
\]
In the one-photon exchange approximation 
its cross section reads  
\begin{equation}
  \D \sigma =
  \frac1{4\ell{\cdot}P} \,
  \frac{e^4}{Q^4} \, L_{\mu\nu} W^{\mu\nu} \, (2\pi)^4 \,
  \frac{\D^3\Vec{\ell}'}{(2\pi)^3 \, 2E'} \,
  \frac{\D^3\Vec{P}_1}{(2\pi)^3 \, 2E_1} \,
  \frac{\D^3\Vec{P}_2}{(2\pi)^3 \, 2E_2} \>, 
  \label{twopart2}
\end{equation}
where $L^{\mu \nu}$ is the ordinary leptonic tensor 
and $W^{\mu \nu}$ is the hadronic tensor:
\bq
W^{\mu \nu} &=& \frac{1}{(2 \pi)^4} \, \sum_a e_a^2 
\, \sum_X \, \int \frac{\D^3 P_X}{(2 \pi)^3 2 E_X} 
\, (2 \pi)^4 \, \delta^4 (q + P - P_X - P_1 - P_2) 
\nonumber \\
& & \times \, \langle P, S \vert J^{\mu}(0) 
\vert P_1, P_2; X \rangle \langle P_1, P_2 ; X \vert 
J^{\nu}(0) \vert P, S \rangle. 
\label{two_hadtens}
\eq

\begin{figure}[t]
\begin{center}
\includegraphics[width=0.60\textwidth]
{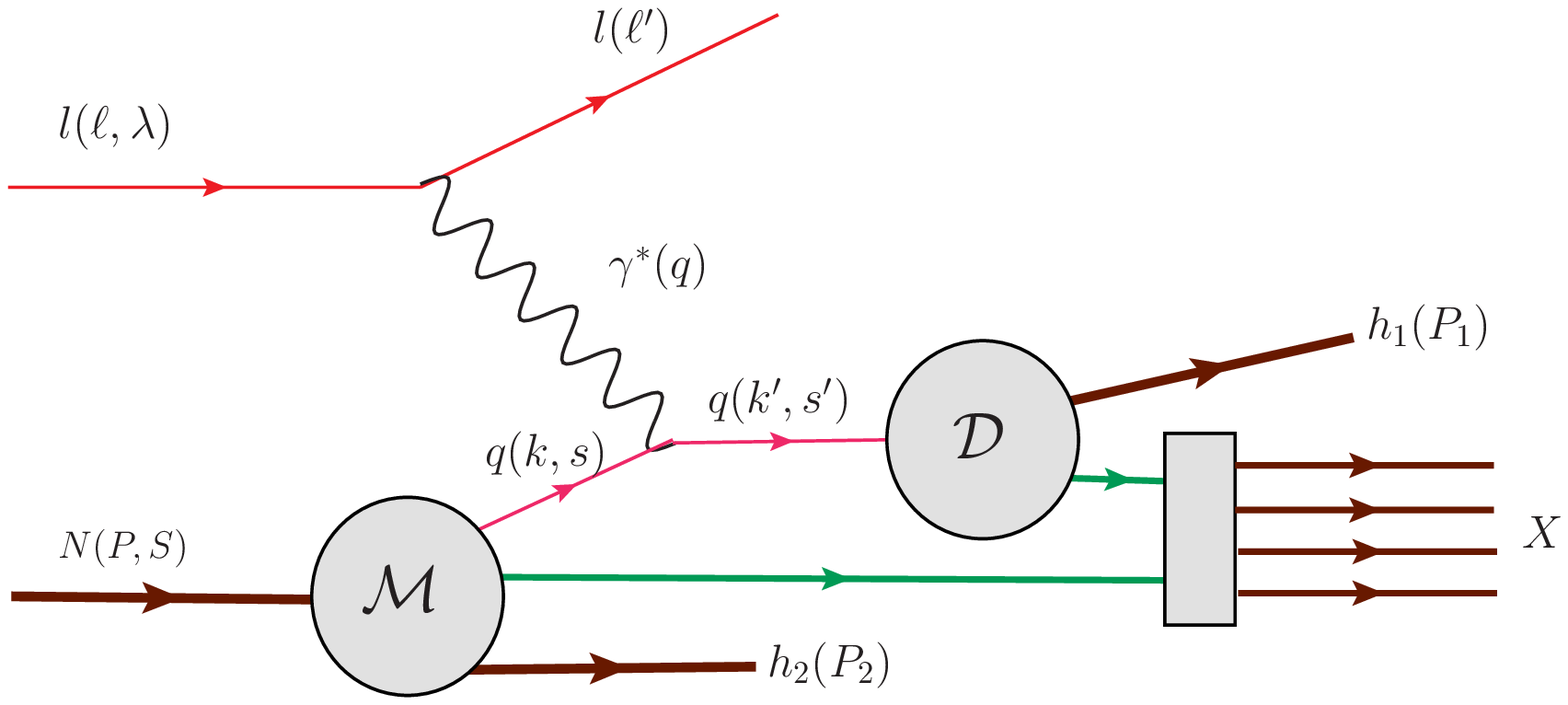}
\caption{Lepto-production
of two hadrons, one in the CFR and one in the TFR.}
\label{sidis_two}
\end{center}
\end{figure}

We are interested in two-particle inclusive lepto-production with 
one hadron ($h_1$) in the CFR and one hadron ($h_2$) in the TFR, 
as shown in Fig.~\ref{sidis_two}. We assume that both hadrons are 
unpolarized or spinless. 

Semi-inclusive DIS is usually described in terms of the three variables 
\begin{equation}
  x_B = \frac{Q^2}{2P{\cdot}q} \quad \quad\quad
  y = \frac{P{\cdot}q}{P{\cdot}\ell} \quad \quad\quad
  z_1 = \frac{P{\cdot}P_1}{P{\cdot}q} \>\cdot
\label{variables}
\end{equation}
When a second hadron, $h_2$, is produced,  
one needs a further variable related to $P_2$. 
It is convenient to use a light-cone parametrization of vectors. 
Given a generic vector $A^{\mu} = (A^0, A^1, A^2, A^2)$,  
their light-cone components are defined as $A^{\pm} \equiv 
(A^0 \pm A^3)/\sqrt{2}$ and we write $A^{\mu} = [A^+, A^-, \Vec A_{\perp}]$. 

We now introduce two null vectors, $n_+^{\mu} = [1, 0, \Vec 0_{\perp}]$ and 
$n_-^{\mu} = [0, 1, \Vec 0_{\perp}]$, 
with $n_+ \cdot n_- = 1$,  so that a vector 
can be parametrized as $A^{\mu} = A^+ n_+^{\mu} 
+ A^- n_-^{\mu} + A_{\perp}^{\mu}$. 
We work in a frame where the target nucleon and the virtual photon 
are collinear (we call it a ``$\gamma^* N$ collinear frame''). 
The nucleon is supposed to move along the $-z$ direction. 

The unit vector $\hat{\Vec q} \equiv \Vec q /\vert \Vec q \vert$ 
identifies the positive $z$ direction. In terms of the null vectors 
$n_+^{\mu}$ and $n_-^{\mu}$ the four-momenta at hand are (approximate 
equalities are valid up to terms proportional to some mass or transverse 
momentum squared):
\bq
& &  P^{\mu} = P^- n_-^{\mu} + \frac{m_N^2}{2 P^-} \, n_+^{\mu}
\simeq P^- n_-^{\mu} \>,  \\
& &  q^{\mu} \simeq \frac{Q^2}{2 x_B P^-} \, n_+^{\mu} - x_B \, P^- n_-^{\mu} 
\>, \\
& & P_1^{\mu}  \simeq \frac{z_1 Q^2}{2 x_B P^-} \, 
n_+^{\mu} + \frac{(\Vec P_{1 \perp}^2 + m_1^2) x_B P^-}{z_1 Q^2} 
\, n_-^{\mu} + P_{1 \perp}^{\mu} 
\simeq \frac{z_1 Q^2}{2 x_B P^-} \, n_+^{\mu} + P_{1 \perp}^{\mu}
\>, \\
& & P_2^{\mu} = \zeta_2 \, P^+ n_+^{\mu} + \frac{\Vec P_{2 \perp}^2 + m_2^2}{2 \zeta P^+} \, n_-^{\mu}
 + P_{2 \perp}^{\mu} \simeq \zeta_2 \, P^+ n_+^{\mu} + P_{2 \perp}^{\mu} \>.
\eq
%


Replacing $\Vec P_1$ and $\Vec P_2$ with the variables 
$(z_1, \Vec P_{1 \perp})$ and $(\zeta_2, \Vec P_{2 \perp})$
respectively, 
the cross section takes the form
\be
  \frac{\D \sigma}{\D x_B \, \D y \, \D z_1 \,\D \zeta_2 \, 
 \D^2 \Vec P_{1 \perp} \, \D^2 \Vec P_{2 \perp} \, \D \phi_S}
  =
  \frac{\alpha_{\rm em}^2}{8 \, (2 \pi)^3 \,  Q^4} \, 
  \frac{y}{z_1 \, \zeta_2} \, L_{\mu\nu} W^{\mu\nu} \,.
  \label{sidis9}
\ee
Here $\phi_S$ is the azimuthal angle of the transverse component
of $S^{\mu}$, the nucleon spin vector, parametrized as
\be
S^{\mu} =
S_{\parallel} \, \frac{P^- n_-^{\mu}}{m_N} - S_{\parallel}
\, \frac{ m_N}{2 P^-} \,
n_+^{\mu} + S_{\perp}^{\mu}
\simeq S_{\parallel}\, \frac{P^- n_-^{\mu}}{m_N} + S_{\perp}^{\mu}\,.
\ee

The explicit expression of the symmetric part of the leptonic
tensor in the $\gamma^* N$ collinear frame is~\cite{Mulders:1995dh}
\bq
L^{\mu \nu}_{\rm (s)} &=&
\frac{Q^2}{y^2} \left \{
- 2 \left ( 1 - y + \frac{y^2}{2} \right ) g_{\perp}^{\mu \nu}
 + 4 (1 - y) \left [
\frac{x_B^2 (P^-)^2}{Q^2}\,   n_-^{\mu} n_-^{\nu} +
\frac{Q^2}{4 x_B^2 (P^-)^2}\, n_+^{\mu} n_+^{\nu} +
\frac{1}{2}\, n_-^{\{ \mu} n_+^{\nu \}} \right ] \right.
\nonumber \\
& & + \left. 4 (1 - y) \left (\hat \ell_{\perp}^{\mu}
\hat \ell_{\perp}^{\nu}  + \frac{1}{2} \, g_{\perp}^{\mu \nu}
\right )
 +  2 (2 - y) \sqrt{1 - y}
\left [ \frac{x_B P^-}{Q} \, n_-^{\{ \mu} \hat \ell_{\perp}^{\nu \}}
+ \frac{Q}{2 x_B P^-}  \, n_+^{\{ \mu} \hat \ell_{\perp}^{\nu \}}
\right ] \right \}\,,
\label{symlept}
\eq
where $\ell_{\perp}^{\mu}$ is the transverse component of the incoming
and outgoing lepton momentum ($\hat\ell_{\perp}^{\mu} =
\ell_{\perp}^{\mu} / |\Vec \ell_{\perp}|$),
and $g_{\perp}^{\mu \nu} = g^{\mu \nu} - (n_+^{\mu} n_-^{\nu}
+ n_+^{\nu} n_-^{\mu})$.
The antisymmetric part of the leptonic tensor
reads ($\lambda$ is the helicity of the lepton
and $\epsilon_{\perp}^{\mu \nu} \equiv 
\epsilon^{\mu \nu}_{\>\>\>\>\> \rho \sigma} \, n_-^{\rho} n_+^{\sigma}$)
\bq
L^{\mu \nu}_{\rm (a)} &=& \frac{Q^2}{y^2} \left \{
- \I \, \lambda \, y (2 - y) \, \epsilon_{\perp}^{\mu \nu}
- 2 \I \, \lambda \, y \sqrt{1 - y} \, 
\epsilon^{\mu \nu}_{\>\>\>\>\>\rho \sigma}
\left ( \frac{x_B P_-}{Q} \, n_-^{\rho} - \frac{Q}{2 x_B P^-} \, n_+^{\rho}
\right ) \hat\ell_{\perp \sigma} \right \} \,.
\label{antlept}
\eq

\begin{figure}[t]
\begin{center}
\includegraphics[width=0.50\textwidth]
{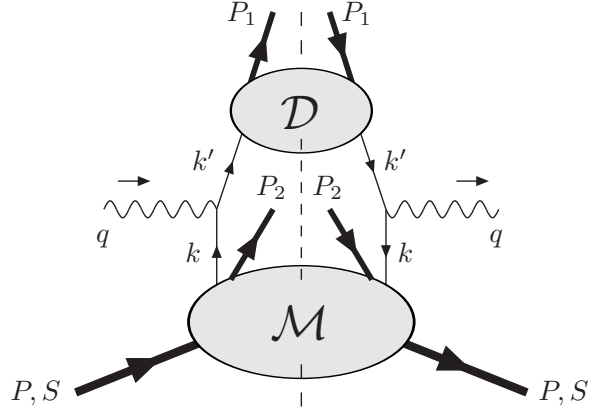}
\caption{The handbag diagram for double hadron lepto-production.}
\label{handbag_two}
\end{center}
\end{figure}

In the parton model, or equivalently at lowest order in QCD,  the
hadronic tensor for the associated 
hadron production in the current and the target 
fragmentation regions  is represented by
the handbag diagram of Fig.~\ref{handbag_two} and reads
(to simplify the presentation, we consider only quarks, the
extension to antiquarks being straightforward):
\bq
W^{\mu \nu} &=& \frac{1}{(2 \pi)^4} \, \sum_a
e_a^2 \, \sum_X \int \frac{\D^3 \Vec P_X}{(2 \pi)^3 \, 2 E_X}
\sum_{X'} \int \frac{\D^3 \Vec P_X'}{(2 \pi)^3 \, 2 E_X'}
\, \int \frac{\D^4 k}{(2 \pi)^4} \,
\int \frac{\D^4 k'}{(2 \pi)^4} \,
\times \nonumber \\
& & \, [ \anti{\chi} (k', P_1) \gamma^{\mu} \phi (k, P, P_2)]^*
[\anti{\chi}(k', P_1) \gamma^{\nu} \phi(k, P, P_2)]
\times \nonumber \\
& &
\, ( 2 \pi)^4 \, \delta^4 (P - k - P_2 - P_X)
\, (2 \pi)^4 \, \delta^4 (k + q - k') \, 
(2 \pi)^4 \, \delta^4 (k' - P_1 - P_X') \,,
\label{hadten0}
\eq
where $\chi$ and $\phi$ are matrix elements of the 
quark field $\psi$ defined as  
\bq
\chi (k', P_1) &=& 
\langle 0 \vert \psi (0) \vert P_1; X' \rangle \>, 
\label{matel1} \\
\phi (k, P, P_2) &=& \langle P_2; X \vert
\psi (0) \vert P, S \rangle \>.  
\label{matel2}
\eq

We now introduce the fracture matrix
$\mathcal{M}$ representing the partonic structure of the nucleon target
when it fragments into the final-state hadron $h_2$:
\bq
 \mathcal{M}_{ij} (k; P,S; P_2) &=&
\sum_X \int \frac{\D^3 \Vec P_X}{(2 \pi)^3 2 E_X}
\int \frac{\D^4 \xi}{(2 \pi)^4} \,
\E^{\I k \cdot \xi} \times \nonumber \\
& &
\langle P, S \vert \anti \psi_j (0) \vert P_2; X \rangle
\langle P_2; X \vert  \psi_i (\xi) \vert P,S \rangle , 
\label{fractmat}
\eq
and the fragmentation matrix $\mathcal{D}$ representing the 
production of the hadron $h_1$ from the current jet: 
\be
 \mathcal{D}_{ij} (k'; P,S; P_1) =
\sum_{X'} \int \frac{\D^3 \Vec P_X'}{(2 \pi)^3 2 E_X'}
\int \frac{\D^4 \eta}{(2 \pi)^4} \,
\E^{\I k' \cdot \eta} \, 
\langle 0 \vert  \psi_i (\eta) \vert P_1; X \rangle
\langle P_1; X \vert  \anti \psi_j (0) \vert 0 \rangle . 
\label{fragmat}
\ee

In QCD, Wilson lines connecting the quark
fields must be inserted  in order to ensure gauge invariance.

Using the definitions (\ref{fractmat}, \ref{fragmat}), the hadronic tensor becomes
\be
W^{\mu \nu} = 
 \sum_a e_a^2 \, \int  \D^4 k \,
\int \D^4 k' 
\, \delta^4 (k + q - k') \, 
 {\rm Tr} \, \left [
\mathcal{M} \gamma^{\mu} \mathcal{D} \gamma^{\nu}
\right ]  \,.
\label{hadten}
\ee

In the parton model description of partly inclusive leptoproduction
it is convenient to use another class of frames,  where the hadron produced in the 
CFR is collinear with the target nucleon (we call them ``$hN$ collinear frames''). 
In these frames the virtual photon acquires a transverse momentum $\Vec q_T^{\mu}$ 
(we use $T$ subscripts to denote transverse quantities in a $hN$ collinear frame). 
The parametrizations of $q^{\mu}$ and  $P_1^{\mu}$ then become
\bq
& &  q^{\mu} \simeq \frac{Q^2}{2 x_B P^-} \, n_+^{\mu} - x_B \, P^- n_-^{\mu} 
+ q_T^{\mu} \>,  
\label{hnparam1} \\
& & P_1^{\mu}  \simeq \frac{z_1 Q^2}{2 x_B P^-} \, 
n_+^{\mu} + \frac{(\Vec P_{1 \perp}^2 + m_1^2) x_B P^-}{z_1 Q^2} 
\, n_-^{\mu}
\simeq \frac{z_1 Q^2}{2 x_B P^-} \, n_+^{\mu} \>.     
\label{hnparam2}
\eq 
One can easily check that the relation between $\Vec q_T$ and $\Vec P_{1 \perp}$ is 
$\Vec q_T = - \Vec P_{1 \perp}/z_1$.   

The quark momenta are parametrized as
\bq
& & k^{\mu} =
x \,P^- n_-^{\mu} + \frac{\Vec k_{T}^2 + k^2}{2 x P^-} \,n_+^{\mu}
+ k_{T}^{\mu} \simeq
x \, P^- n_-^{\mu} + k_{T}^{\mu} \>, 
\label{kparam1} \\
& & 
{k'}^{\mu} = \frac{P_1^+}{z} \, n_+^{\mu} + 
\frac{z ({k'}^2 + {\Vec {k}'}_T^2)}{2 P_1^+} \, n_-^{\mu} + {k'}_T^{\mu} 
\simeq \frac{P_1^+}{z} \, n_+^{\mu} + {k'}_T^{\mu} \>. 
\label{kparam2} 
\eq
Here $x$ is the fraction of the light-cone momentum of the 
target carried by the emitted quark, and $z$ is the 
fraction of the light-cone momentum of the struck quark 
carried by the hadron in the CFR. 
The delta function in Eq.~(\ref{hadten}) enforces the constraints 
$x = x_B$ and $z = z_1$. Written explicitly, the hadronic 
tensor is
\bq
W^{\mu \nu} &=& 
 \sum_a e_a^2 \, \int  \D k^+ \, \D k^- \, \D^2 \Vec k_T \,
\int \D {k'}^+ \, \D {k'}^- \, \D^2 \Vec k_T'  
\nonumber \\
& & \times 
 \delta ({k'}^+ - P_1^+/z_1) \, 
\delta (k^- - x_B P^-) \, \delta^2 (\Vec k_T + \Vec q_T 
- \Vec k_T') \, 
 {\rm Tr} \, \left [
\mathcal{M} \gamma^{\mu} \mathcal{D} \gamma^{\nu}
\right ]. 
\label{hadtenx}
\eq

Notice that while one should in principle distinguish between transverse 
vectors in a $\gamma^* N$ collinear frame (labelled by a $\perp$ subscript) 
and transverse vectors in a $hN$ collinear frame (labelled by a 
$T$ subscript), the difference is of order $(P^-)^2$ and can 
be ignored as far as one neglects subleading corrections 
 in $P^-$ (i.e., higher twists).

\section{Transverse momentum dependent fracture functions and 
fragmentation functions}

The most general decomposition of the fracture matrix $\mathcal{M}$ in a basis 
of Dirac matrices would contain terms proportional to $\one, \gamma^{\mu},
\gamma^{\mu} \gamma_5, \gamma_5, \sigma^{\mu \nu} \gamma_5$.
We are interested in leading-twist fracture functions, i.e. in
terms of $\mathcal{M}$ that are of order $(P^-)^1$.
At this order, only the vector, axial and tensor components
of $\mathcal{M}$ appear~\cite{Barone:2001sp}:
\be
\mathcal{M} = \frac{1}{2} \, ( \mathcal{V}_{\mu} \gamma^{\mu}
+ \mathcal{A}_{\mu}  \gamma_5 \gamma^{\mu}
+ \I \, \mathcal{T}_{\mu \nu}\, \sigma^{\mu \nu} \gamma_5 ) \,,
\label{decmat}
\ee
where the coefficients $\mathcal{V}^{\mu}$, $\mathcal{A}^{\mu}$
and $\mathcal{T}^{\mu \nu}$ contain various combinations of the vectors, 
or pseudo-vectors, $P^{\mu}, P_1^{\mu}, P_2^{\mu}, k^{\mu}, k'^{\mu}$ and $S^{\mu}$.

The polarized transverse-momentum
dependent fracture functions appear in the expansion of the leading twist
Dirac projections ($\Gamma = \gamma^-,\gamma^- \gamma_5, \I \sigma^{i-} \gamma_5$)
\bq
& & \mathcal{M}^{[\Gamma]} (x_B, \Vec k_{\perp},
\zeta_2, \Vec P_{2 \perp})
\nonumber \\
& & \hspace{1cm}
\equiv
\frac{1}{4 \zeta_2} \int \frac{\D k^+ \, \D k^-}{(2 \pi)^3}
\, \delta (k^- - x_B P^-) \, {\rm Tr} \, (\mathcal{M} \, \Gamma)
\nonumber \\
& &
\hspace{1cm}
=
\frac{1}{4 \zeta_2}
\, \int \frac{\D \xi^+ \,
\D^2 \Vec \xi_{\perp}}{(2 \pi)^6}
\, \E^{\I (x_B P^- \xi^+ - \Vec k_{\perp}
\cdot \Vec \xi_{\perp})} \,
\sum_X \int \frac{\D^3 \Vec P_X}{(2 \pi)^3 \, 2 E_X}
\times \nonumber \\
& & \hspace{1.5cm}
\langle P, S\vert \anti \psi(0) \Gamma
\vert  P_2; X \rangle \langle  P_2; X \vert
 \psi(\xi^+, 0, \Vec \xi_{\perp})
\vert P, S \rangle \,.
\label{proj}
\eq
These represent the conditional probabilities to find an unpolarized
($\Gamma = \gamma^-$),  a longitudinally polarized
($\Gamma = \gamma^- \gamma_5$) or a transversely polarized
($\Gamma = \I \sigma^{i-} \gamma_5$) quark with longitudinal momentum
fraction $x_B$ and transverse momentum $\Vec k_{\perp}$ inside a nucleon
fragmenting into a hadron carrying a fraction $\zeta_2$ of the nucleon
longitudinal momentum and a transverse momentum $\Vec P_{2 \perp}$.
Again, in QCD a Wilson line $\mathcal{W}$ must be inserted, which
for $\Vec k_{\perp}$-dependent distributions includes transverse links
and is generally rather complicated \cite{Belitsky:2002sm,Bomhof:2006dp}:
its explicit structure, however, is irrelevant for our purposes.

The most general parameterization of the traced fracture matrix 
(\ref{proj}) at leading twist is:
\bq
\mathcal{M}^{[\gamma^-]}
&=&
\hat{u}_1
+ \frac{\Vec P_{2 \perp} \times \Vec S_{\perp}}{m_2} \,
\hat{u}_{1T}^h + \frac{\Vec k_{\perp} \times
\Vec S_{\perp}}{m_N} \, \hat{u}_{1T}^{\perp}
 +
\frac{S_{\parallel} \, (\Vec k_{\perp}
\times \Vec P_{2 \perp})}{m_N \, m_2} \,
\hat{u}_{1L}^{\perp h}
\label{v1v2} \\
\mathcal{M}^{[\gamma^- \gamma_5]}
&=&
S_{\parallel} \, \hat{l}_{1L}
+ \frac{ \Vec P_{2 \perp} \cdot \Vec S_{\perp}}{m_2} \,
 \hat{l}_{1T}^h
+ \frac{\Vec k_{\perp} \cdot \Vec S_{\perp}}{m_N}
\,  \hat{l}_{1T}^{\perp}
 +  \frac{\Vec k_{\perp} \times
\Vec P_{2 \perp}}{m_N \, m_2} \,
 \hat{l}_1^{\perp h}
\label{a1a2} \\
\mathcal{M}^{[\I \, \sigma^{i -} \gamma_5]}
&=& S_{\perp}^i \,  \hat{t}_{1T}
+ \frac{S_{\parallel} \, P_{2 \perp}^i}{m_2} \,  \hat{t}_{1L}^h
+ \frac{S_{\parallel} \, k_{\perp}^i}{m_N} \,
 \hat{t}_{1L}^{\perp}
\nonumber \\
& & + \, \frac{(\Vec P_{2 \perp} \cdot \Vec S_{\perp})
\, P_{2 \perp}^i}{m_2^2} \,  \hat{t}_{1T}^{hh}
+ \frac{(\Vec k_{\perp} \cdot \Vec S_{\perp})
\, k_{\perp}^i}{m_N^2} \,  \hat{t}_{1T}^{\perp \perp}
\nonumber \\
& & + \, \frac{(\Vec k_{\perp} \cdot \Vec S_{\perp})
\, P_{2 \perp}^i - (\Vec P_{2 \perp} \cdot \Vec S_{\perp})
\, k_{\perp}^i }{m_N m_2} \,  \hat{t}_{1T}^{\perp h}
\nonumber \\
& & + \, \frac{\epsilon_{\perp}^{ij} P_{2 \perp j}}{m_2}
\,  \hat{t}_1^h
+ \frac{\epsilon_{\perp}^{ij} k_{\perp j}}{m_N}
\,  \hat{t}_1^{\perp}\,,
\label{t1t2}
\eq
where by the vector product of the two-dimensional vectors we mean the
pseudo-scalar quantity $\Vec a_{\perp} \times \Vec b_{\perp} =
\epsilon_{\perp i j} \, a_{\perp}^i b_{\perp}^ j = \vert
\Vec a _{\perp} \vert \vert \Vec b_{\perp} \vert \, \sin (\phi_b - \phi_a)$.
All fracture functions depend on the scalar variables
$x_B, \Vec k_{\perp}^2, \zeta_2, \Vec P_{2 \perp}^2,
\Vec k_{\perp} \cdot \Vec P_{2 \perp}$. 

Notice that, with respect to Ref.~\cite{Anselmino:2011ss}, 
we have adopted a new, Amsterdam--style, nomenclature 
for fracture functions, which makes their correspondence with 
distribution functions more visible. In particular, we denote by 
$\hat{u}$ (formerly, $\hat{M}$) 
the unintegrated fracture functions of unpolarized 
quarks, by $\hat{l}$ (formerly, $\Delta \hat{M}$) the unintegrated 
fracture functions of longitudinally polarized quarks, and 
by $\hat{t}$ (formerly, $\Delta_T \hat{M}$) the unintegrated 
fracture functions of transversely polarized quarks. 
The subscript 1 denotes leading-twist quantities. 
The subscripts $L$ and $T$ label the polarization of the target 
(no subscript = unpolarized, $L$ = longitudinally polarized, $T$ = transversely 
polarized). The superscripts $h$ and $\perp$ signal the presence of factors 
$P_{2 \perp}^i$ and $k_{\perp}^i$, respectively. Fracture functions 
integrated over the hadron transverse momentum will have no hat; fracture 
functions integrated over the quark transverse momentum will have a tilde. 

An important point to stress is that while parity invariance constrains the 
structure of the fracture matrix, time reversal invariance does not, since
$\mathcal{M}$, similarly to the fragmentation matrix, contains the 
out-states $\vert P_2; X \rangle$.

Turning now to the fragmentation matrix, its Dirac projections are defined 
as
\bq
 \mathcal{D}^{[\Gamma]} (z_1, \Vec k_{\perp}') &\equiv& 
\frac{1}{4 z_1} \, \int \D k'^+ \int \D k'^- 
\, \delta(k'^+ - P_1^+/z_1) \, {\rm Tr} \, (\mathcal{D} \Gamma)
\nonumber \\
&=& 
\frac{1}{4 z_1} \, 
 \int \frac{\D \eta^- \, \D^2 \Vec \eta_{\perp}}{(2 \pi)^3} \, 
\E^{\I (P_1^+ \eta^-/z_1 - \Vec k'_{\perp} 
\cdot \Vec \eta_{\perp}} \, 
\sum_{X'} \int \frac{\D^3 \Vec P_X'}{(2 \pi)^3 2 E_X'}
 \times \nonumber \\
& &
\langle 0 \vert \Gamma \psi_i (0, \eta^-, \Vec \eta_{\perp}) \vert P_1; X \rangle
\langle P_1; X \vert  \anti \psi_j (0) \vert 0 \rangle , 
\label{fragproj}
\eq

At leading twist, and considering unpolarized or spinless hadrons, there 
are only two fragmentation functions: 
\bq
 \mathcal{D}^{[\gamma^+]} &=& D_1 \>, 
\label{d1} \\
 \mathcal{D}^{[\I \sigma^{i +} \gamma_5]} 
&=& \frac{\epsilon_{\perp}^{ij} \, k'_{\perp j}}{m_1} \, H_1^{\perp}\,. 
\label{h1collins}
\eq
Here $D_1$ is the ordinary unpolarized fragmentation function, whereas 
$H_1^{\perp}$ is the Collins function, describing the fragmentation of 
transversely polarized quarks into an unpolarized hadron.  


\section{The hadronic tensor} 

Using the Fierz decomposition 
\bq
(\gamma^{\mu})_{ij} \, (\gamma^{\nu})_{kl} 
&=& 
\frac{1}{4} \, 
\left \{ g^{\mu \nu} \, \left [ 
- (\gamma_{\rho})_{il} \, (\gamma^{\rho})_{kj} 
- (\gamma_{\rho} \gamma_5)_{il} (\gamma^{\rho} \gamma_5 )_{kj} 
+ \frac{1}{2} \, (\I \sigma_{\alpha \beta} \gamma_5)_{il} 
\, (\I \sigma^{\alpha \beta} \gamma_5)_{kj} \right ] \right. \nonumber \\
& & + (\gamma^{\{ \mu})_{il} \, (\gamma^{\nu \} })_{kj} + 
(\gamma^{\{ \mu} \gamma_5)_{il} \, (\gamma^{\nu \} } \gamma_5)_{kj}
- (\I \sigma^{\{ \mu}_{\alpha} \gamma_5)_{il} \, 
(\I \sigma^{\nu \} \alpha} \gamma_5 )_{kj} 
\nonumber \\
& & + \left. \Strut \epsilon^{\mu \nu \rho \sigma} 
\, \left [ (\gamma_{\rho})_{il} \,  (\gamma_{\sigma} \gamma_5)_{kj} 
+ (\gamma_{\rho} \gamma_5 )_{il} \, (\gamma_{\sigma})_{kj} 
\right ] \right \} + \ldots,  
\label{fierz} 
\eq
where the dots label terms that do not contribute to leading twist, we can 
re-express the hadronic tensor (\ref{hadtenx}) as (we retain the leading-twist contributions only and return to the $\gamma^* N$ collinear frame)
\bq
W^{\mu \nu} &=& 4 z_1 \zeta_2 \, (2 \pi)^3 \, \sum_a e_a^2 
\, \int \D^2 \Vec k_{\perp} \, \int \D^2 \Vec k_{\perp}' \, 
\delta^2 (\Vec k_{\perp} - \Vec k_{\perp}' - \Vec P_{1 \perp}/z_1) 
\nonumber \\
& & \times \left \{ - g_{\perp}^{\mu \nu} \, 
\mathcal{M}^{[\gamma^-]} \mathcal{D}^{[\gamma^+]} 
+ g_{\perp}^{\mu \nu} \, 
\mathcal{M}^{[\I \sigma_i^- \gamma_5]} \mathcal{D}^{[\I \sigma^{i +} \gamma_5]}
- \mathcal{M}^{[\I \sigma^{\{ \mu -} \gamma_5]} \mathcal{D}^{[\I \sigma^{\nu \} +} \gamma_5]}
\right.
\nonumber \\
& & \left. 
+ \I \, \epsilon_{\perp}^{\mu \nu} \,  
\mathcal{M}^{[\gamma^- \gamma_5]} \mathcal{D}^{[\gamma^+]}
\right \}. 
\label{hadten2} 
\eq
The first term in Eq.~(\ref{hadten2}) couples the unpolarized fracture 
functions to the unpolarized fragmentation function $D_1$. 
The second and third terms involve the transversely polarized fracture 
functions and the Collins fragmentation function $H_1^{\perp}$.     
The last term represents the antisymmetric part of $W^{\mu \nu}$, which 
contributes to lepto-production with a polarized beam and couples the 
longitudinally polarized fracture functions to $D_1$. 

Using Eqs.~(\ref{v1v2}-\ref{t1t2}) and  
Eqs.~(\ref{d1}-\ref{h1collins}), we get 
\bq
W^{\mu \nu} &=& 4 z_1 \zeta_2 \, (2 \pi)^3 \, \sum_a e_a^2 
\, \int \D^2 \Vec k_{\perp} \, \int \D^2 \Vec k_{\perp}' \, 
\delta^2 (\Vec k_{\perp} - \Vec k_{\perp}' - \Vec P_{1 \perp}/z_1) 
\nonumber \\ 
& & \times 
\left \{ - g_{\perp}^{\mu \nu} \, 
\left [ \hat{u}_1 \, D_1 + \frac{\Vec P_{2 \perp} \times \Vec S_{\perp}}{m_2} 
\, \hat{u}_{1T}^h \, D_1 + \frac{\Vec k_{\perp} \times \Vec S_{\perp}}{m_N} 
\, \hat{u}_{1T}^{\perp} \, D_1 + 
\frac{  S_{\parallel} (\Vec k_{\perp} \times \Vec P_{2 \perp})}{m_N m_2} 
\, \hat{u}_{1L}^{\perp h} \, D_1 \right ] \right. 
\nonumber \\
& & 
-  \frac{\left (S_{\perp}^{\{ \mu} \epsilon_{\perp}^{\nu \} \rho} k'_{\perp \rho} 
+ {k'}_{\perp}^{\{ \mu} \epsilon_{\perp}^{\nu \} \rho} S_{\perp \rho} 
\right )}{2 m_1} \,  \hat{t}_{1T} \, H_1^{\perp} 
\nonumber \\
& & 
- S_{\parallel} \,  
\frac{\left (P_{2 \perp}^{\{ \mu} \epsilon_{\perp}^{\nu \} \rho} k'_{\perp \rho} 
+ {k'}_{\perp}^{\{ \mu} \epsilon_{\perp}^{\nu \} \rho} P_{2 \perp \rho} 
\right )}{2 m_1 m_2} \,  \hat{t}_{1L}^h \, H_1^{\perp} 
\nonumber \\
& & 
 - S_{\parallel} \, 
\frac{\left (k_{\perp}^{\{ \mu} \epsilon_{\perp}^{\nu \} \rho} k'_{\perp \rho} 
+ {k'}_{\perp}^{\{ \mu} \epsilon_{\perp}^{\nu \} \rho} k_{\perp \rho} 
\right )}{2 m_1 m_N} \,  \hat{t}_{1L}^{\perp} \, H_1^{\perp} 
\nonumber \\
& & 
+ \frac{P_{2 \perp} \cdot S_{\perp} \, 
\left (P_{2 \perp}^{\{ \mu} \epsilon_{\perp}^{\nu \} \rho} k'_{\perp \rho} 
+ {k'}_{\perp}^{\{ \mu} \epsilon_{\perp}^{\nu \} \rho} P_{2 \perp \rho} 
\right )}{2 m_1 m_2^2} \,  \hat{t}_{1T}^{hh} \, H_1^{\perp} 
\nonumber \\
& & 
+ \frac{k_{\perp} \cdot S_{\perp} \, 
\left (k_{\perp}^{\{ \mu} \epsilon_{\perp}^{\nu \} \rho} k'_{\perp \rho} 
+ {k'}_{\perp}^{\{ \mu} \epsilon_{\perp}^{\nu \} \rho} k_{\perp \rho} 
\right )}{2 m_1 m_N^2} \,  \hat{t}_{1T}^{\perp \perp} \, H_1^{\perp} 
\nonumber \\
& & 
+  \frac{k_{\perp} \cdot S_{\perp} \, 
\left (P_{2 \perp}^{\{ \mu} \epsilon_{\perp}^{\nu \} \rho} k'_{\perp \rho} 
+ {k'}_{\perp}^{\{ \mu} \epsilon_{\perp}^{\nu \} \rho} P_{2 \perp \rho} 
\right ) 
+ P_{2 \perp} \cdot S_{\perp}  
\left (k_{\perp}^{\{ \mu} \epsilon_{\perp}^{\nu \} \rho} k'_{\perp \rho} 
+ {k'}_{\perp}^{\{ \mu} \epsilon_{\perp}^{\nu \} \rho} k_{\perp \rho} 
\right )}{2 m_1 m_2 m_N} \,  \hat{t}_{1T}^{\perp h} \, H_1^{\perp} 
\nonumber \\
& & 
+ 
\frac{P_{2 \perp}^{\{ \mu} {k'}_{\perp}^{\nu \}} - 
g_{\perp}^{\mu \nu}\, P_{2 \perp} \cdot k'_{\perp}}{m_1 m_2} 
\,  \hat{t}_1^h \, H_1^{\perp} 
+ \frac{k_{\perp}^{\{ \mu} {k'}_{\perp}^{\nu \}} - 
g_{\perp}^{\mu \nu}\, k_{\perp} \cdot k'_{\perp}}{m_1 m_N} 
\,  \hat{t}_1^{\perp} \, H_1^{\perp} 
\nonumber \\ 
& &  + \I \, \epsilon_{\perp}^{\mu \nu} \, 
\left [ S_{\parallel} \,  \hat{l}_{1L} \, D_1 
+ \frac{\Vec P_{2 \perp} \cdot \Vec S_{\perp}}{m_2} 
\,  \hat{l}_{1T}^h \, D_1 
\right. 
\nonumber \\
& & + \left. \left. \frac{\Vec k_{\perp} \cdot \Vec S_{\perp}}{m_N} 
\,  \hat{l}_{1T}^{\perp} \, D_1 + 
\frac{ \Vec k_{\perp} \times \Vec P_{2 \perp}}{m_N m_2} 
\,  \hat{l}_1^{\perp h} \, D_1  \right ] 
\right \}. 
\label{hadten_full}
\eq 

The fully differential cross section for two-hadron production 
is obtained by contracting the hadronic tensor (\ref{hadten_full})  
with the leptonic tensor, Eqs.~(\ref{symlept}, \ref{antlept}). 
The final expression is extremely complicated and will be reported elsewhere. 
In the following we will focus on double lepto-production integrated 
over the transverse momentum either of the hadron produced 
in the TFR, or of the hadron produced in the CFR.   

\section{Double hadron lepto-production integrated over $\Vec P_{2 \perp}$}

If we integrate the fracture matrix over $\Vec P_{2 \perp}$ 
we are left with eight $k_{\perp}$-dependent fracture functions:  
\bq
\int \D^2 \Vec P_{2 \perp} \, \mathcal{M}^{[\gamma^-]}
&=&
u_1 + \frac{\Vec k_{\perp} \times
\Vec S_{\perp}}{m_N} \, u_{1T}^{\perp} \>, 
\label{v1v2_tilde} \\
\int \D^2 \Vec P_{2 \perp} \, 
\mathcal{M}^{[\gamma^- \gamma_5]}
&=&
S_{\parallel} \,  l_{1L}
+ \frac{\Vec k_{\perp} \cdot \Vec S_{\perp}}{m_N}
\,  l_{1T} \>, 
\label{a1a2_tilde} \\
\int \D^2 \Vec P_{2 \perp} \, \mathcal{M}^{[\I \, \sigma^{i -} \gamma_5]}
&=& S_{\perp}^i \,  t_{1T}
+ \frac{S_{\parallel} \, k_{\perp}^i}{m_N} \,
 t_{1L}^{\perp}
 + \frac{k_{\perp}^i (\Vec k_{\perp} \cdot \Vec S_{\perp})}{m_N^2} 
\,   t_{1T}^{\perp}  
+ \frac{\epsilon_{\perp}^{ij} k_{\perp j}}{m_N}
\,  t_1^{\perp}
\nonumber \\
&=& S_{\perp}^i \,  t_1
+ \frac{S_{\parallel} \, k_{\perp}^i}{m_N} \,
 t_{1L}^{\perp}
+ \frac{(k_{\perp}^i k_{\perp}^j - \frac{1}{2} \Vec k_{\perp}^2 \delta_{ij}) \, 
S_{\perp}^j}{m_N^2} \,  t_{1T}^{\perp}  
+ \frac{\epsilon_{\perp}^{ij} k_{\perp j}}{m_N}
\,  t_1^{\perp}\,,
\label{t1t2_tilde}
\eq
where $ t_{1} \equiv t_{1T} 
+ (\Vec k_{\perp}^2/2 m_N^2) \,  t_{1T}^{\perp}$. 
We have removed the hat to denote the $\Vec P_{2 \perp}$--integrated 
fracture functions: 
\bq
& & u_1 (x_B, \Vec k_{\perp}^2, \zeta_2) = \int \D^2 \Vec P_{2 \perp} \,
\hat{u}_1 \> , 
\label{mtilde1} \\
& & u_{1T}^{\perp} (x_B, \Vec k_{\perp}^2, \zeta_2) = 
 \int \D^2 \Vec P_{2 \perp}
\left \{
\hat{u}_{1T}^{\perp} + \frac{m_N}{m_2}
\, \frac{\Vec k_{\perp} \cdot \Vec P_{2 \perp}}{\Vec
k_{\perp}^2} \, \hat{u}_{1T}^h \right \},
\label{mtilde2}  \\
& &  l_{1L} (x_B, \Vec k_{\perp}^2, \zeta_2) = 
 \int \D^2 \Vec P_{2 \perp}
 \hat{l}_{1L} \>, 
\label{mtilde3} \\
& &  l_{1T} (x_B, \Vec k_{\perp}^2, \zeta_2) = 
 \int \D^2 \Vec P_{2 \perp}
\left \{  \hat{l}_{1T}^{\perp} +
\frac{m_N}{m_2} \, \frac{\Vec k_{\perp} \cdot \Vec P_{2 \perp}}{\Vec
k_{\perp}^2} \, \hat{l}_{1T}^h \right \}  , 
\label{mtilde4} \\
& &  t_1 (x_B, \Vec k_{\perp}^2, \zeta_2) = 
 \int \D^2 \Vec P_{2 \perp}
\left \{  \hat{t}_{1T} + \frac{\Vec k_{\perp}^2}{2m_N^2} \,
 \hat{t}_{1T}^{\perp\perp} + \frac{\Vec P_{2\perp}^2}{2m_2^2} \,
 \hat{t}_{1T}^{hh} \right \}.  
\label{mtilde8} \\
& &  t_{1L}^{\perp} (x_B, \Vec k_{\perp}^2, \zeta_2) = 
 \int \D^2 \Vec P_{2 \perp}
\left \{  \hat{t}_{1L}^{\perp} +
\frac{m_N}{m_2} \, \frac{\Vec k_{\perp} \cdot \Vec P_{2 \perp}}{\Vec
k_{\perp}^2} \,  \hat{t}_{1L}^h \right \} ,  
\label{mtilde5} \\
& & t_{1T}^{\perp} (x_B, \Vec k_{\perp}^2, \zeta_2) = 
 \int \D^2 \Vec P_{2 \perp}
\left \{  \hat{t}_{1T}^{\perp \perp} 
+ \frac{m_N^2}{m_2^2}\frac{
2 (\Vec k_{\perp} \cdot \Vec P_{2 \perp})^2
- \Vec k_{\perp}^2 \Vec P_{2 \perp}^2}{(\Vec k_{\perp}^2)^2}
 \,  \hat{t}_{1T}^{hh}
\right \} , 
\label{mtilde7} \\
& &  t_1^{\perp} (x_B, \Vec k_{\perp}^2, \zeta_2) = 
 \int \D^2 \Vec P_{2 \perp}
\left \{  \hat{t}_1^{\perp} +
\frac{m_N}{m_2} \, \frac{\Vec k_{\perp} \cdot \Vec P_{2 \perp}}{\Vec
k_{\perp}^2} \,  \hat{t}_1^h \right \}. 
\label{mtilde6} 
\eq
By virtue of the sum rules derived in Ref.~\cite{Anselmino:2011ss}, 
these fracture functions are directly related to the eight leading--twist  distribution functions by 
\bq
& & \sum_h \int \D \zeta_2 \, \zeta_2 \, 
u_1 (x_B, \Vec k_{\perp}^2, \zeta_2) 
 = (1 - x_B) \, f_1 (x_B, \Vec k_{\perp}^2) \>, 
\label{f1} \\
& & \sum_h \int \D \zeta_2 \, \zeta_2 \, 
u_{1T}^{\perp} (x_B, \Vec k_{\perp}^2, \zeta_2) 
= - (1 - x_B) \, f_{1T}^{\perp} (x_B, \Vec k_{\perp}^2) \>, 
\label{f1T}  \\
& & \sum_h \int \D \zeta_2 \, \zeta_2 \,
 l_{1L} (x_B, \Vec k_{\perp}^2, \zeta_2) 
= (1 - x_B) \, g_{1L} (x_B, \Vec k_{\perp}^2) \>,
\label{g1L} \\
& & \sum_h \int \D \zeta_2 \, \zeta_2 \, 
 l_{1T} (x_B, \Vec k_{\perp}^2, \zeta_2)  =
(1 - x_B) \, g_{1T}(x_B, \Vec k_{\perp}^2) \>,
\label{g1T} \\
& &
\sum_h \int \D \zeta_2 \, \zeta_2 \, 
 t_1 (x_B, \Vec k_{\perp}^2, \zeta_2) 
= (1 - x_B) \, h_{1}(x_B, \Vec k_{\perp}^2) \>,
\label{h1} \\
& & \sum_h \int \D \zeta_2 \, \zeta_2 \,
 t_{1L}^{\perp} (x_B, \Vec k_{\perp}^2, \zeta_2) 
 =
(1 - x_B) \, h_{1L}^{\perp}(x_B, \Vec k_{\perp}^2) \>, 
\label{h1L} \\
& &
\sum_h \int \D \zeta_2 \, \zeta_2 \, 
 t_{1T}^{\perp} (x_B, \Vec k_{\perp}^2, \zeta_2) 
 =
(1 - x_B) \, h_{1T}^{\perp}(x_B, \Vec k_{\perp}^2) \>, 
\label{h1Tperp} \\
& &
\sum_h \int \D \zeta_2 \, \zeta_2 \, 
 t_1^{\perp} (x_B, \Vec k_{\perp}^2, \zeta_2) 
= - (1 - x_B) \, h_{1}^{\perp}(x_B, \Vec k_{\perp}^2) \>.  
\label{h1perp} 
\eq
Notice that, among the 16 fracture functions listed in 
Eqs.~(\ref{v1v2}-\ref{t1t2}), the three functions with double superscript 
$\perp \! h$, i.e. $\hat{u}_{1L}^{\perp h}$, $\hat{l}_1^{\perp h}$ and 
$\hat{h}_{1T}^{\perp h}$, which measure correlations 
involving both the quark and the hadron transverse momenta,  
have no distribution function counterpart and disappear once the integration 
over any of the two transverse momenta is performed. In particular, $\hat{l}_{1}^{\perp h}$, which describes longitudinally polarized quarks 
inside an unpolarized nucleon, is not probed in single hadron 
lepto-production, whereas it gives rise to a  beam spin asymmetry in 
(unintegrated) two--hadron lepto-production \cite{Anselmino:2011bis, 
Kotzinian:2011fb}.   
 
The $\Vec P_{2 \perp}$--integrated hadronic tensor reads 
\bq
\int \D^2 \Vec P_{2 \perp} \, W^{\mu \nu} &=& 4 z_1 \zeta_2 \, (2 \pi)^3 \, \sum_a e_a^2 
\, \int \D^2 \Vec k_{\perp} \, \int \D^2 \Vec k_{\perp}' \, 
\delta^2 (\Vec k_{\perp} - \Vec k_{\perp}' - \Vec P_{1 \perp}/z_1) 
\nonumber \\ 
& & \times 
\left \{ - g_{\perp}^{\mu \nu} \, 
\left [ u_1 \, D_1 +  \frac{\Vec k_{\perp} \times \Vec S_{\perp}}{m_N} 
\, u_{1T}^{\perp} \, D_1  \right ] \right. 
\nonumber \\
& & 
-  \frac{\left (S_{\perp}^{\{ \mu} \epsilon_{\perp}^{\nu \} \rho} k'_{\perp \rho} 
+ {k'}_{\perp}^{\{ \mu} \epsilon_{\perp}^{\nu \} \rho} S_{\perp \rho} 
\right )}{2 m_1} \,  t_{1T} \, H_1^{\perp} 
\nonumber \\
& & 
 - S_{\parallel} \, 
\frac{\left (k_{\perp}^{\{ \mu} \epsilon_{\perp}^{\nu \} \rho} k'_{\perp \rho} 
+ {k'}_{\perp}^{\{ \mu} \epsilon_{\perp}^{\nu \} \rho} k_{\perp \rho} 
\right )}{2 m_1 m_N} \,  t_{1L}^{\perp} \, H_1^{\perp} 
\nonumber \\
& & 
+ \frac{k_{\perp} \cdot S_{\perp} \, 
\left (k_{\perp}^{\{ \mu} \epsilon_{\perp}^{\nu \} \rho} k'_{\perp \rho} 
+ {k'}_{\perp}^{\{ \mu} \epsilon_{\perp}^{\nu \} \rho} k_{\perp \rho} 
\right )}{2 m_1 m_N^2} \,  t_{1T}^{\perp} \, H_1^{\perp} 
\nonumber \\
& & 
+ \frac{k_{\perp}^{\{ \mu} {k'}_{\perp}^{\nu \}} - 
g_{\perp}^{\mu \nu}\, k_{\perp} \cdot k'_{\perp}}{m_1 m_N} 
\,  t_1^{\perp} \, H_1^{\perp} 
\nonumber \\ 
& &  + \I \, \left. \epsilon_{\perp}^{\mu \nu} \, 
\left [ S_{\parallel} \,  l_{1L} \, D_1 + 
 \frac{\Vec k_{\perp} \cdot \Vec S_{\perp}}{m_N} 
\,  l_{1T} \, D_1   \right ]  
\right \}. 
\label{hadten_integr}
\eq 
This hadronic tensor is perfectly analogous 
to the one describing single-hadron leptoproduction 
in the CFR \cite{Mulders:1995dh}, the correspondence being:  
Fracture Functions $\Rightarrow$ Distribution Functions. 
Thus we can use the procedure of Ref.~\cite{Mulders:1995dh} to contract 
the hadronic tensor and the leptonic tensor. The final expression 
of the cross section is 
 \bq
& & 
 \frac{\D \sigma}{\D x_B \, \D y \, \D z_1 \, \D \zeta_2 \,     
\D \phi_1 \, \D  P_{1\perp}^2 \, \D \phi_S} = 
 \frac{\alpha_{\rm em}^2}{ x_B \, y \, Q^2} \left \{ 
\left (1 - y + \frac{y^2}{2} \right ) \, \mathcal{F}_{UU, T} 
+ (1 - y) \, \cos 2 \phi_1 \, \mathcal{F}_{UU}^{\cos 2 \phi_1}  \right. 
\nonumber \\
& & \hspace{2cm} + \, S_{\parallel} \, 
 (1 - y) \, \sin 2 \phi_1 \, \mathcal{F}_{UL}^{\sin 2 \phi_1}
 + S_{\parallel} \, \lambda_{\ell} 
\,  y \, \left (1 - \frac{y}{2} \right ) \, \mathcal{F}_{LL}
\nonumber \\
& & \hspace{2cm} + \, S_{\perp} \, 
 \left (1 - y + \frac{y^2}{2} \right ) \, \sin (\phi_1 - \phi_S) \, 
\mathcal{F}_{UT}^{\sin (\phi_1 - \phi_S)}  
\nonumber \\
& & \hspace{2cm} + \, S_{\perp} \,  (1 -y) \, \sin (\phi_1 + \phi_S) \, \mathcal{F}_{UT}^{\sin (\phi_1 + 
\phi_S)} + S_{\perp} \, (1 - y ) \, \sin (3 \phi_1 - \phi_s) \, 
\mathcal{F}_{UT}^{\sin (3 \phi_1 - \phi_S)} 
\nonumber \\
& & \hspace{2cm} + \left.  S_{\perp} \,  \lambda_{\ell} 
\,  y \left (1 - \frac{y}{2} \right ) \, \cos (\phi_1 - \phi_S) \, 
\mathcal{F}_{LT}^{\cos (\phi_1 - \phi_S)} 
\right \}  
\label{sidiscs_lt}
\eq
where the structure functions are given at leading twist by  
($\hat{\Vec P}_1 \equiv \Vec P_{1 \perp}/\vert \Vec P_{1 \perp} \vert$)
\bq
\mathcal{F}_{UU,T} &= & \mathcal{C} \, \left [u_1 D_1 \right ] , 
\label{sf1} \\
\mathcal{F}_{UU}^{\cos 2 \phi_1} &=& 
\mathcal{C} 
\left [ \frac{2 (\hat{\Vec P}_1 \cdot \Vec k_{\perp}) (\hat{\Vec P}_1 
\cdot \Vec k_{\perp}') - \Vec k_{\perp} \cdot \Vec k_{\perp}'}{m_N m_1}  
\,  t_1^{\perp} \, H_1^{\perp} \right ] , 
\label{sf3} \\
\mathcal{F}_{UL}^{\sin 2 \phi_1} &=& 
\mathcal{C} 
\left [ -\frac{2 (\hat{\Vec P}_1 \cdot \Vec k_{\perp}) (\hat{\Vec P}_1 
\cdot \Vec k_{\perp}') - \Vec k_{\perp} \cdot \Vec k_{\perp}'}{m_N m_1} 
\,  t_{1L}^{\perp} \, H_1^{\perp} \right ] , 
\label{sf4} \\
\mathcal{F}_{LL} &=& \mathcal{C} \, \left [  l_{1L} D_1  \right ],  
\label{sf5} \\
\mathcal{F}_{UT}^{\sin (\phi_1 - \phi_S)} &=& 
\mathcal{C} \left [ \frac{\hat{\Vec P}_1 \cdot \Vec k_{\perp}}{m_N} \, 
u_{1T}^{\perp} D_1 \right ] , 
\label{sf6} \\
\mathcal{F}_{UT}^{\sin (\phi_1 + \phi_S)} &=& 
\mathcal{C} \left [ - \frac{\hat{\Vec P}_1 \cdot \Vec k_{\perp}'}{m_1} \, 
 t_1 H_1^{\perp} \right ] , 
\label{sf8} \\
\mathcal{F}_{UT}^{\sin (3 \phi_1 - \phi_S)} &=& 
\mathcal{C} 
\left [ \frac{ 2 (\hat{\Vec P}_1 \cdot \Vec k_{\perp}') (\Vec k_{\perp} 
\cdot \Vec k_{\perp}') + \Vec k_{\perp}^2 (\hat{\Vec P}_1 \cdot \Vec k_{\perp}') - 
4 (\hat{\Vec P}_1 \cdot \Vec k_{\perp})^2 (\hat{\Vec P}_1 \cdot \Vec k_{\perp}')}{2 
m_N^2 m_1} \,  t_{1T}^{\perp} H_1^{\perp} \right ] , 
\label{sf9} \\
\mathcal{F}_{LT}^{\cos (\phi_1 - \phi_S)} &=& 
\mathcal{C} \left [ \frac{\hat{\Vec P}_1 \cdot \Vec k_{\perp}}{m_N} \, 
 l_{1T} D_1 \right ],  
\label{sf9bis}
\eq 
with the following notation 
for the transverse momenta convolutions
\bq
\mathcal{C} \, [w u D] &=& 
 \sum_a e_a^2 \, x_B \, \int \D^2 \Vec k_{\perp} \int \D^2 \Vec k_{\perp}'
\, \delta^2 ( \Vec k_{\perp} - \Vec k_{\perp}' - \Vec P_{1 \perp}/z_1) 
\nonumber \\
& & \times \>
 w(\Vec k_{\perp}, \Vec k_{\perp}') 
\, u^a (x_B, \Vec k_{\perp}^2, \zeta_2) \, D^a (z_1, \Vec k_{\perp}^{'2})\,.  
\label{convol}
\eq

\section{Double hadron lepto-production integrated over $\Vec P_{1 \perp}$}

If one integrates the hadronic tensor over the transverse momentum 
$\Vec P_{1 \perp}$ of the hadron produced in the CFR, the two integrations 
over $\Vec k_{\perp}$ and $\Vec k_{\perp}'$ in Eq.~(\ref{hadten2}) 
disentangle and can be performed separately. 

Integrating the fragmentation matrix over $\Vec k_{\perp}'$, only one 
fragmentation function, $D_1$, survives, which couples to the unpolarized 
and the longitudinally polarized fracture functions.  
The relevant fracture matrix projections integrated over $\Vec k_{\perp}$ are \cite{Anselmino:2011ss} 
\bq
 \int \D^2 \Vec k_{\perp} \, \mathcal{M}^{[\gamma^-]}
&=& \tilde{u}_1 (x_B, \zeta_2, \Vec P_{2 \perp}^2)
+ \frac{\Vec P_{2 \perp} \times \Vec S_{\perp}}{m_2}
\, \tilde{u}_{1T}^h (x_B, \zeta_2, \Vec P_{2 \perp}^2), 
\label{intkfrag1} \\
 \int \D^2 \Vec k_{\perp} \,
\mathcal{M}^{[\gamma^- \gamma_5]}
&=& S_{\parallel} \,  \tilde{l}_{1L} (x_B, \zeta_2, \Vec P_{2 \perp}^2)
+ \frac{\Vec P_{2 \perp} \cdot \Vec S_{\perp}}{m_2}
\, \tilde{l}_{1T}^h (x_B, \zeta_2, \Vec P_{2 \perp}^2),
\label{intkfrag2}
\eq
with (a tilde denotes fracture functions integrated 
over the quark transverse momentum)
\bq
& & \tilde{u}_1(x_B, \zeta_2, \Vec P_{2 \perp}^2)
= \int \D^2 \Vec k_{\perp} \,
\hat{u}_1 , 
\label{intkfrag3} \\
& & \tilde{u}_{1T}^h (x_B, \zeta_2, \Vec P_{2 \perp}^2)
= \int \D^2 \Vec k_{\perp} \,
\left \{ \hat{u}_{1T}^h
+ \frac{m_2}{m_N}
\frac{\Vec k_{\perp} \cdot \Vec P_{2 \perp}}{\Vec P_{2 \perp}^2}
\, \hat{u}_{1T}^{\perp} \right \}, 
\label{intkfrag4} \\
& & \tilde{l}_{1L} (x_B, \zeta_2, \Vec P_{2 \perp}^2)
= \int \D^2 \Vec k_{\perp} \,
 \hat{l}_{1L} ,
\label{intkfrag5} \\
& & \tilde{l}_{1T}^h (x_B, \zeta_2, \Vec P_{2 \perp}^2)
= \int \D^2 \Vec k_{\perp}
\, \left \{  \hat{l}_{1T}^h +
\frac{m_2}{m_N}
\frac{\Vec k_{\perp} \cdot \Vec P_{2 \perp}}{\Vec P_{2 \perp}^2}
\, \hat{l}_{1T}^{\perp} \right \}. 
\label{intkfrag6} 
\eq
The final result for the cross section is 
\bq
& &  \frac{\D \sigma}{\D x_B \, \D y \, \D z_1 \, \D \zeta_2 \,    
\D \phi_2 \, \D  P_{2 \perp}^2 \, \D \phi_S} = 
 \frac{\alpha_{\rm em}^2}{y \, Q^2} \, 
\left \{ \left (1 - y + \frac{y^2}{2} \right ) \right. 
\nonumber \\
& & \hspace{1cm}
\times \, \sum_a e_a^2 \,
 \left  [   \tilde{u}_1(x_B, \zeta_2, \Vec P_{2 \perp}^2) 
-   \vert \Vec S_{\perp} \vert \, \frac{\vert \Vec P_{2 \perp}\vert}{m_2}
\, \tilde{u}_{1T}^h (x_B, \zeta_2, \Vec P_{2 \perp}^2) \, \sin (\phi_2 - \phi_S)
\right ]
\nonumber \\
& &  \hspace{1cm} + \,
\lambda \, y \, \left (1 - \frac{y}{2}\right )
\sum_a e_a^2 \,
 \left [ \STRUT
S_{\parallel} \, \tilde{l}_{1L} (x_B, \zeta_2, \Vec P_{2 \perp}^2)
\right.
\nonumber \\
& & \hspace{1cm}
+ \, \left. \left.
\vert \Vec S_{\perp} \vert \, \frac{\vert \Vec P_{2 \perp} \vert}{m_2}
\,  \tilde{l}_{1T}^h (x_B, \zeta_2, \Vec P_{2 \perp}^2) \, \cos (\phi_2 - \phi_S)
\right ] \right \} D_1 (z_1) . 
\label{crossintk}
\eq
As in the case of single-hadron production \cite{Anselmino:2011ss}, there 
is a Sivers-type modulation $\sin (\phi_2 - \phi_S)$, but no Collins-type
effect.    

\section{Conclusions and perspectives}

We have considered the double production of unpolarized hadrons in deep 
inelastic scattering processes, with one hadron produced in the current 
fragmentation region and one in the target fragmentation region. We have 
combined the fracture function formalism, which describes the hadron 
production in the TFR \cite{Trentadue:1993ka,Grazzini:1997ih} 
and the fragmentation function formalism, which describes the hadron 
production in the CFR. Target polarization and the transverse motions, 
of quarks inside the parent hadron, and in the fragmentation process, 
have been taken into account. TMD factorization has been assumed. 

This papers completes a previous one \cite{Anselmino:2011ss} in which 
the formalism of polarized and transverse momentum dependent fracture 
functions was introduced to describe single hadron production in the TFR. 
The combined observation of one hadron in the TFR and one hadron in the 
CFR allows access to a class of chiral-odd fracture functions which cannot 
contribute to single hadron production. 

The general result for the hadronic tensor involving all fracture and 
fragmentation functions is given in Eq.~(\ref{hadten_full}), and the 
corresponding cross section can be obtained by inserting this result into 
Eq.~(\ref{sidis9}) and using Eqs.~(\ref{symlept}) and (\ref{antlept}).
We have not presented explicitly the final expression, which is rather 
cumbersome, but have considered more realistic cases in which 
one only measures the longitudinal component of one of the final hadrons, integrating over its transverse momentum. Such results are given in 
Eqs.~(\ref{sidiscs_lt}) and (\ref{crossintk}).         
      
Eq.~(\ref{sidiscs_lt}), which refers to the case in which one detects the 
three momentum $(z_1, \Vec P_{1 \perp})$ of one hadron in the CFR (like in 
the usual SIDIS) and the longitudinal momentum fraction $\zeta_2$ of 
another hadron in the TFR, has the same structure as the familiar cross 
section for single hadron production in the CFR, with the role of the 
distribution functions (TMDs) replaced by the fracture functions. 
Therefore, it has the same potentiality, for measuring the 
$\Vec P_{2 \perp}$-integrated fracture functions, as the usual CFR SIDIS 
for measuring the TMDs. 

Eq.~(\ref{crossintk}), which refers to the case in which one detects the 
three momentum $(\zeta_2, \Vec P_{2 \perp})$ of one hadron in the TFR and 
the longitudinal momentum fraction $z_1$ of another hadron in the CFR, 
has the same structure as the cross section for the single hadron 
production in the TFR, Eq.~(51) of Ref.~\cite{Anselmino:2011ss}, with the 
addition of an integrated fragmentation function. The presence of these 
fragmentation functions in principle allows to perform the quark flavor 
decomposition of quark transverse momentum integrated fracture functions 
as it is done for distribution functions in SIDIS in the CFR.

Phenomenological analyses of SIDIS data, based on the results presented 
here and in Ref.~\cite{Anselmino:2011ss}, could confirm our full 
understanding of the mechanism of hadron production in lepton-nucleon
interactions. The observation and measurement of the predicted azimuthal dependences in the TFR would allow the extraction of the fracture functions, 
similarly to what is being done for TMDs in the CFR. 

The clear disentanglement of effects observed in the two regions, 
TFR and CFR, is crucial for an unambiguous interpretation of the data. 
Although some first results might be available soon from JLab and COMPASS,
we think that the ideal experiments to test the fracture function 
factorization and measure these new functions in SIDIS, are those being 
discussed in the international community and planned at future Electron Ion 
or Electron Nucleon Colliders (EIC/ENC). 

\section*{Acknowledgements}
We acknowledge partial support by the Italian Ministry of Education,
University and Research (MIUR) in the framework of a Research Project
of National Interest (PRIN 2008).
We also acknowledge partial support by the European Community - Research
Infrastructure Activity under the FP7 program (HadronPhysics2, Grant
agreement 227431), by the Helmholtz Association through
funds provided to the Virtual Institute ``Spin and Strong QCD''(VH-VI-231)
and by Regione Piemonte.



\begin{thebibliography}{99}

\bibitem{Anselmino:2011ss} 
M.~Anselmino, V.~Barone and A.~Kotzinian, {\it Phys. Lett.} 
{\bf B699} (2011) 108, arXiv:1102.4214 [hep-ph]. 

\bibitem{Mulders:1995dh}
P.J.~Mulders and R.D.~Tangerman, {\it Nucl. Phys.} {\bf B461} (1996) 197,
hep-ph/9510301.

\bibitem{Barone:2001sp}
V.~Barone, A.~Drago and P.G.~Ratcliffe, {\it Phys. Rep.} {\bf 359} (2002) 1,
hep-ph/0104283.

\bibitem{Belitsky:2002sm}
A.V.~Belitsky, X.~Ji and F.~Yuan,
{\it Nucl. Phys.} {\bf B656} (2003) 165, hep-ph/0208038.

\bibitem{Bomhof:2006dp}
C.J.~Bomhof, P.J.~Mulders and F.~Pijlman,
{\it Eur. Phys. J.} {\bf C47} (2006) 147,
hep-ph/0601171.

\bibitem{Anselmino:2011bis} 
M.~Anselmino, V.~Barone and A.~Kotzinian, paper in preparation. 

\bibitem{Kotzinian:2011fb} 
A.~Kotzinian, M.~Anselmino and V.~Barone, arXiv:1107.2292 [hep-ph]. 

\bibitem{Trentadue:1993ka}
L.~Trentadue and G.~Veneziano, {\it Phys. Lett.} {\bf B323} (1994) 201.






\bibitem{Grazzini:1997ih}
M.~Grazzini, L.~Trentadue and G.~Veneziano,
{\it Nucl. Phys.} {\bf B519} (1998) 394, hep-ph/9709452.

























\end{thebibliography}
\end{document}